\documentclass[useAMS,usenatbib]{mn2e}
\usepackage[dvips]{graphicx,color}
\usepackage{txfonts}
\usepackage{amssymb}
\usepackage{indentfirst}
\usepackage{epsfig}

\newcommand{\MgXII}{\mbox{{Mg\,{\sevensize XII}}}}

\newcommand{\FeSi}{\ensuremath{\mbox{Fe~K}\alpha/\mbox{Si~K}\alpha~}}

\newcommand{\Lyb}{\ensuremath{\hbox{Ly}\beta~}}
\newcommand{\Ka}{\ensuremath{\hbox{K}\alpha~}}

\newcommand{\A}{\AA~}

\def\cha{{\it Chandra }}

\begin{document}
\title[]{X-ray signatures of the polar dusty gas in AGN}

\author[J. Liu et al.]{Jiren Liu$^{1,2}$\thanks{E-mail: jirenliu@nao.cas.cn},
	Sebastian F. H$\ddot{\rm o}$nig$^{2}$, Claudio Ricci$^{3,4}$, and St\'{e}phane Paltani$^{5}$\\ 
	 $^{1}$National Astronomical Observatories, 20A Datun Road, Beijing 100012, China\\
	 $^{2}$Department of Physics and Astronomy, University of Southampton, Southampton SO17 1BJ, UK\\
	 $^{3}$N\'{u}cleo de Astronom\'{\i}a  de la Facultad de Ingenier\'{\i}a Universidad Diego Portales, 
	 Av. Ej\'{e}rcito Libertador 441, Santiago, Chile \\
	 $^{4}$Kavli Institute for Astronomy and Astrophysics, Peking University, Beijing 100871, China\\
	 $^{5}$Department of Astronomy, University of Geneva, 1205 Versoix, Switzerland\\
}

\date{}

\maketitle

\begin{abstract}

Recent mid-infrared interferometry observations of nearby active galactic nuclei (AGN) 
revealed that a significant part of the dust emission 
extends in the polar direction, rather than the equatorial torus/disk direction as expected
by the traditional unification model.
We study the X-ray signatures of this polar dusty gas with ray-tracing simulations.
Different from those from the ionized gas, the scattered emission from the polar dusty gas 
produces self-absorption and neutral-like fluorescence lines, which are potentially a unique 
probe of the kinematics of the polar dusty gas.
The anomalously small \FeSi ratios of type II AGN observed previously can be 
naturally explained by the polar dusty gas, because the polar emission does not suffer 
from heavy absorption by the dense equatorial gas.
The observed Si \Ka lines of the Circinus galaxy and NGC 1068 show blue-shifts with respect to 
the systemic velocities of the host galaxies, consistent with an outflowing scenario of the 
Si K$\alpha$-emitting gas.
The 2.5-3 keV image of the Circinus galaxy is elongated along the polar direction, consistent 
with an origin of the polar gas. These results show that the polar-gas-scattered X-ray emission 
of type II AGN is an ideal objective for future X-ray missions, such as Athena.

\end{abstract}


\section{Introduction}

The traditional unification model of active galactic nuclei (AGN) assumes an axisymmetric 
dusty torus that obscures the accretion disk and the broad-line region when the system 
is observed edge-on \citep[e.g.][]{Ant93}. 
Recent mid-infrared (IR) interferometry observations of nearby AGN revealed that 
a significant part of the dust emission 
extends in the polar direction, instead of the equatorial torus direction
\cite[for a recent review, see][]{Bur16}. 
The Circinus galaxy and NGC 1068 are the two best 
studied sources, for which multiple components are dissected \citep{Tri14,LG14}.
This polar dust component could be the inner wall of a torus \citep{Tri14}, or a dusty wind driven
by radiation pressure \citep{Hon12} or by magnetocentrifugal force \citep{Vol18}. For 
the wind scenario, the polar dust could be distributed 
along the walls of a hollow cone \citep{Hon13, Sta19}.
Single-dish mid-IR images of local AGN show ubiquitous evidences of extended emission 
on tens to hundreds pc
scales along the polar direction \citep[e.g.][and references therein]{Asm16}.
The polar dust component is also evidenced by its effect on IR SED \citep[e.g.][]{Lyu18}.
The optical depth of the polar dust can not be too large, otherwise the emission 
produced in the ionization cone would be strongly depleted. 
In some sources \citep[e.g. NGC 1068,][]{Ant94}, the optical polarization was observed to be wavelength independent
on $1''$ scale, implying that the electron scattering is the dominant scattering mechanism.
It indicates that the polar dust distribution on such scales might be clumpy.

In this paper we study X-ray signatures of this polar dusty gas, 
which might be observable for obscured/type II AGN, especially for Compton-thick 
sources, where 
the direct X-ray emission of AGN is heavily obscured \citep[e.g.][]{Com04}. In fact, 
both the Circinus galaxy and NGC 1068 are Compton-thick.

When X-ray photons interact with matter, they can be absorbed or scattered. Photons absorbed
by inner-shell ionization can further lead to the emission of fluorescence
photons. The scattered X-ray spectrum is characterized by a Compton hump (around 10-30 keV) and 
a series of fluorescent lines, with Fe \Ka (6.4 keV) being the most prominent one, due to the 
high abundance and high fluorescence yield of iron.
As a result, the fluorescence lines can be used to probe the geometrical distribution and kinematics
of the gas surrounding the X-ray sources.
Many calculations of scattered X-ray emission have been done assuming an accretion disk or a torus 
\citep[e.g.][]{GF91,MZ95,MY09,Liu14,Tan19}.
While the torus/disk is likely being Compton-thick, the polar dusty gas has a relatively 
low column density. Therefore, the polar gas will mainly affect the low-energy 
X-ray emission. Most importantly, the emission of the polar gas in obscured AGN 
can reach the observer directly, without being obscured by the dense equatorial gas.
As a result, the polar dusty gas can produce much stronger low-energy emission 
in edge-on viewing angles compared with the equatorial gas. While the Fe \Ka lines 
have been extensively studied for toridal geometries \citep[e.g.][]{Ric14}, the X-ray signatures
of a polar dusty gas are rarely studied and compared with observations.
\citet{Liu16} measured the fluorescence line ratios of \FeSi for a sample of nearby obscured 
AGN and found that they are an order of magnitude lower than
those predicted by clumpy torus models. In other words, the observed Si \Ka lines 
are much stronger than what would be expected by the equatorial torus gas.
In principle, the polar dusty gas could explain these anomalous fluorescence line ratios. 

The importance of scattered X-ray emission from the ionized wind
in type II AGN (which shown as warm absorbers in type I AGN) was recognized 
from early days \citep[e.g.][]{KK95}. The ionized winds of AGN have been extensively studied
\citep[e.g.][]{Pro00}, with recent studies also involving dusty winds 
\citep[e.g.][]{Dor16,CK17,Wil19}. The existence of dust within warm absorbers has been reported
\citep[e.g.][]{Lee01,MC18}. The polar dusty gas could be the dusty and densest part
of the dusty winds.
Different from those from the ionized wind, the scattered emission from the polar dusty wind
will be absorbed by the dusty wind itself, and neutral-like fluorescence lines will be produced. 
The fluorescence characteristic of the dusty wind is of special interests, because
the X-ray fluorescence lines are potentially a powerful probe of the kinematics of the dusty wind,
which is crucial to tell the physical nature and possible effects of the dusty wind. It is generally 
hard to obtain the kinematic information of dusty gas from infrared observations, 
as thermal dust emission lacks sharp line features.

\section{Gas model and simulation method}

\begin{figure}
\includegraphics[width=2.8in]{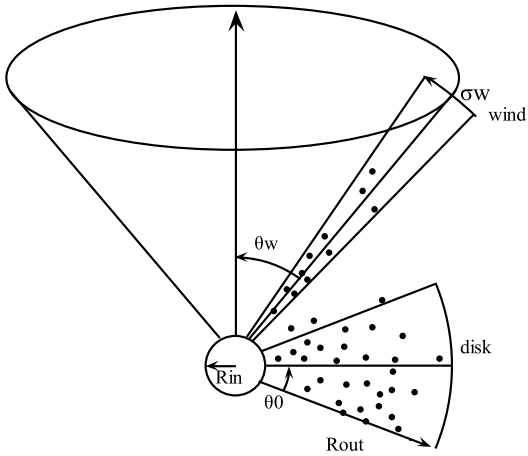}
\caption{Illustration of the simulated geometry of the equatorial disk and the polar wind 
	(hollow cone).
}
\end{figure}

For the equatorial gas we use the generally adopted clumpy model \citep[e.g.][]{Nen08a,HK10}.
Since the equatorial component is likely to be a disk \citep[e.g. the Circinus
galaxy was observed to have a disk-like emission with a size of $0.2\times1.1$ pc
by][]{Tri14}, we 
refer to it as "disk" below.
The clumps are distributed according to a radial power law ($\propto r^a$, within a region of 
$R_{in}$ and $R_{out}$) and a Gaussian distribution in the elevation direction 
($\propto {\rm exp}[-(\pi/2-\theta)^2/\theta_0^2$]), where $\theta$ is the half-opening angle and 
$\theta_0$ is the angular extent of the disk (Figure 1).
The total number of clumps ($N_{tot}$) is determined 
by $N_0$, the mean number of clumps along a radial path in the equatorial plane, and the radius of 
clumps $R_c$. Finally, the density of individual clumps ($n_c$) completes the description.
$R_c$ and $n_c$ can be combined as the average column density of a clump 
$N_{Hc}$ ($\sim1.33R_cn_c$).

The inner boundary $R_{in}$ is determined by the distance at which the dust sublimates for a given
AGN luminosity. We adopt $R_{in}=0.1$\,pc, corresponding to a luminosity 
$\sim5\times10^{44}$ erg\,s$^{-1}$ \citep{Kis07}. $R_{out}/R_{in}$ is taken to be 10, as inferred from
the clumpy modeling of infrared observations \citep{Nen08b}.
The density distribution index $a\approx-3$ if the gravitational potential is dominated by a black hole and
$a\approx-1$ for a stellar cluster \citep[e.g.][]{BD04}. We set $a=-1$, consistent with that inferred 
by \citet{Nen08b}. We take $\theta_0=\pi/8$ and $N_0=7$, the typical values inferred by \citet{Nen08b}.
The clump column density $N_{Hc}$ is set to be $4.4\times10^{23}$\,cm$^{-2}$, providing a Compton-thick 
obscuration ($N_H=3\times10^{24}$\,cm$^{-2}$) along the radial path in the equatorial plane.
The sky covering factor of the angular extent of 
$\theta_0=\pi/8$ is about 0.4. The total simulated clump number is 37000 
(with $R_c=0.0074$\,pc), and the volume filling factor in the equatorial plane is
$0.06$ ($\sim\frac{N_0R_c}{R_{out}-R_{in}}$). 
Reflected X-ray emission of clumpy tori 
have been studied previously \citep[e.g.][]{Liu14,Tan19}. \citet{Liu14} found that the volume filling 
factor only slightly influences the reflected spectra, while the total column density 
and the mean number of clumps ($N_0$) can significantly affect the reflected emission.
This is because the sky covering factor of a clumpy torus is 
$\propto1-e^{-N_0}$. That is, the reflected emission is not restricted to the 
absolute size scale and only sensitive to the total column density, the mean number of clumps, 
and the angular distribution.

Following \citet{HK17}, we model the polar gas component (referred as "wind" below) 
as a hollow cone, which is characterized by 
a radial distribution ($\propto r^{a_w}$, within $R_{in}^w$ and $R_{out}^w$),
an half opening angle ($\theta_w$), and an angular 
width $\sigma_w$. Similar to the equatorial clumps, the clumps in the hollow cone are described by
the mean number of clumps along a radial path $N_{0}^w$ and the average clump 
column density $N_{Hc}^{w}$.
Since the observed polar dust emission is more extended than the equatorial one 
\citep[e.g. the polar component of the Circinus
galaxy has a size of $0.8\times1.9$ pc inferred by][]{Tri14}, 
we adopt a larger region 
of $R_{out}^w=2$\,pc for the polar wind. 
The other parameters are set as $R_{in}^w=0.1$\,pc, $a_w=0$,
$\theta_w=\pi/4$, $\sigma_w=\pi/12$, $N_0^w=2.5$, and 
$N_{Hc}^{w}=1.5\times10^{22}$\,cm$^{-2}$,
corresponding to a wind column density of 
$3.8\times10^{22}$\,cm$^{-2}$ along a radial path.
The total number of clumps in the hollow cone is 12000 (with $R_c^w=0.0148$\,pc).
An illustration of the simulated geometry is shown in Figure 1.
We note that the adopted configuration is a general representation of 
a compact disk plus a polar wind, not corresponding to a real system.
Generally the observed mid-infrared interferometry radii of AGN are 10-20 times larger 
than the dust sublimation radii measured from near-IR dust reverberation mapping \citep{Bur13}.
If the spatial extent of the wind is much larger than that of the disk, the scattered emission 
by the wind would not depend much on the presence of the disk. Thus, we also simulate the situation
of a wind and a disk alone.
The dependences of the scattered emission on different parameters are studied in \S 3.2.

We use RefleX code developed by \citet{PR17} to
perform the simulations. RefleX is a ray-tracing code designed for the study
of propagation of X-ray photons in the matter surrounding an X-ray source.
RefleX can model a large variety of matter geometries, which are defined by
the user using an arbitrary number of simple geometrical building blocks.
RefleX implements a photon-by-photon Monte Carlo simulation: Each X-ray
photon is emitted following a user-specified spectral distribution and a
user-specified geometrical distribution; it is then propagated through the
surrounding medium, undergoing repeatedly all the usual processes that take
place in the X-ray domain, namely Compton and Rayleigh scattering,
photoelectric absorption and fluorescence emission, until the photon is
either destroyed, or escapes the simulated object, in which case it can be
used to build spectra or images. RefleX has been validated by comparison
with existing models, like pexmon \citep{MZ95} or MYTorus \citep{MY09}.
We refer to \citet{PR17} for more
details. The simulation presented here assume a power-law input spectrum
with a photon index of 1.8 in the $1-100$ keV energy range \citep[e.g.][]{NP94}.
All atoms are assumed to be neutral in the simulation. The gas
composition is assumed to be the Solar abundances from \citet{AG89}.

\section{Simulation Results}

\begin{figure*}
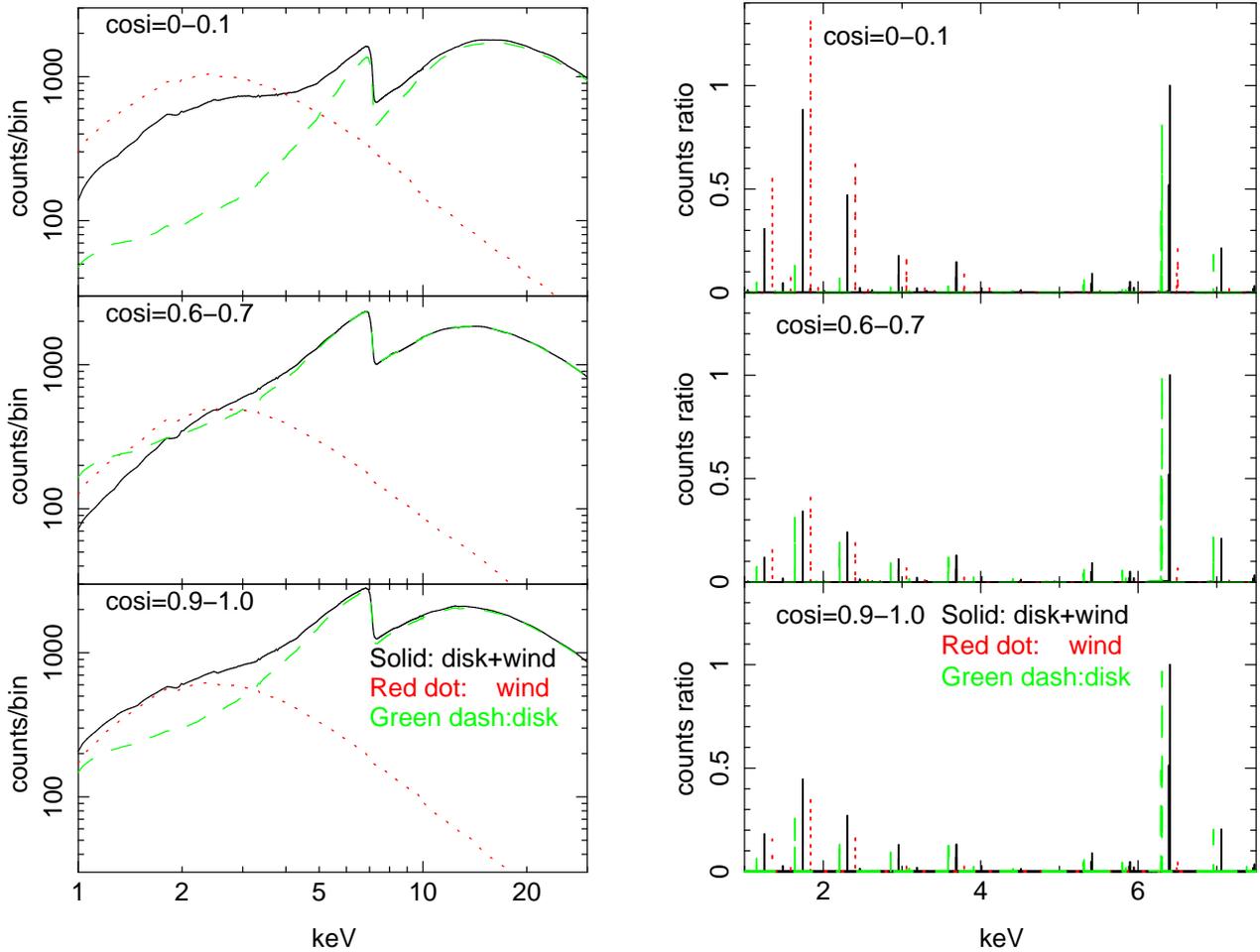

\hspace*{-0.5cm}
\includegraphics[width=3.5in]{H_NH324sca.ps}
\includegraphics[width=3.5in]{H_NH324fluo.ps}
\caption{Scattered X-ray continuum (left) and fluorescence lines (right) for 
	models of disk+wind (black solid lines), disk (green dash lines), and wind (red dot lines) at 
	different inclination angles. The fluorescence lines are normalized 
	by the Fe \Ka counts of the disk+wind model. For viewing purpose,
	the lines below 6 keV are increased by a factor of 10, and those 
	of the disk and the wind model are left and right
	shifted by 0.1 keV, respectively. 
}
\end{figure*}

\subsection{Spectral results}

As noted in the introduction, the scattered X-ray emission by a dusty wind would have different 
characteristics from that by an ionized wind, as a dusty wind produces self-absorption and neutral-like
fluorescence lines.
To study the scattered X-ray signature of the polar dusty wind,
we have calculated the scattered X-ray emission for geometries of disk, wind, and disk+wind, separately.
In this section we present the results of the default setting listed in \S 2, and their dependences 
on adopted parameters are studied in \S 3.2.
The resulting continuum and fluorescence lines at different inclination angles for different geometries
are plotted in the left and right panels of Figure 2, respectively.

The most notable difference between the disk+wind model and the disk model occurs for 
inclination angles around 90 degrees, i.e., viewing the system edge-on.
As can be seen from the left panel of Figure 2, for $\cos i=0-0.1$, the emission below 
5 keV is 5-10 times stronger in the disk+wind model than in the disk model.
The reason is that the emission of the polar wind above the disk can reach the 
observer directly, suffering no heavy absorption by the dense equatorial gas.
The emission of the wind model below 5 keV is about twice that of the 
disk+wind model. This is because in the disk+wind model, the emission of the 
inner part of the wind component is absorbed by the disk component.
On the other hand, the high energy end of the scattered emission of the disk+wind model 
is dominated by the disk component. 
This is expected, since the scattering of high-energy photons is more effective 
for large column densities.
The self-absorption of the wind is seen as the drop of the scattered continuum of the disk 
model below 2 keV.

For $\cos i=0.6-0.7$ (similar to the direction of the hollow cone $\theta_w=\pi/4$),
the spectrum of the disk+wind model 
is similar to that of the disk model, except at the lowest energy ($\sim1$ keV), where
the spectrum of the disk+wind model is a little weaker than that of the disk model. This 
is due to the extra absorption provided by the wind component.

\begin{figure}
\includegraphics[width=3.4in]{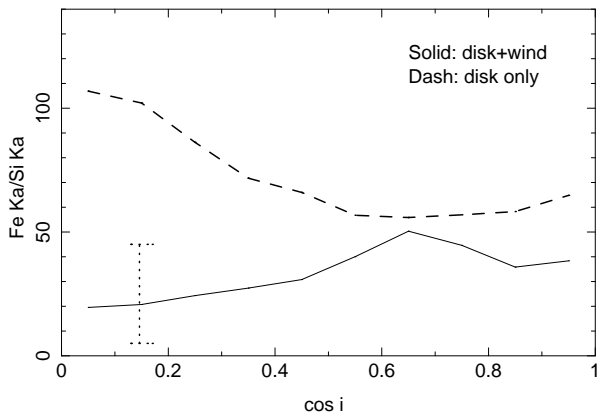}
\caption{\FeSi ratios predicted by the disk+wind and disk models. The vertical dotted lines around 
	$\cos i=0.15$ indicate the observed ranges ($5-45$) measured for type II AGN, all of which 
	are heavily obscured \citep{Liu16}.
}
\end{figure}

For $\cos i=0.9-1.0$, the face-on viewing angles, the flux of the disk+wind model spectrum 
is also higher than
that of the disk model spectrum, but only with a factor of 2. Compared with the case of 
$\cos i=0.6-0.7$, the contribution of the wind component is a little higher. It is because 
for face-on viewing angles, the absorption by the wind component is less effective.

The corresponding fluorescence lines (right panel of Figure 2) show a similar trend as 
the scattered continuum.
For example, the Si \Ka photons in edge-on angles of the disk+wind model are 6 
times more abundant than those of the disk model, and most of the Si \Ka photons
are produced by the wind component. Note that $\sim30$\% Si \Ka photons of the wind 
component are absorbed by the disk component, while the Fe \Ka photons of the 
disk+wind model are close to the sum of the wind and the disk components.

The line ratios of \FeSi for the disk and disk+wind models are plotted in Figure 3,
along with the observed ranges ($5-45$) measured from nearby type II AGN \citep{Liu16}.
As can be seen, in edge-on viewing angles, the \FeSi ratio of the disk model is 
5-10 times larger than
the observed values, while the \FeSi ratio of the disk+wind model is about 20,
consistent with the observed values. The dependences of the results on different
parameters are studied in next section.

\subsection{Spectral results: parameter study}

Since the scattered emission of AGN looks prominent only for type II AGN, 
the intrinsic continuum of which are heavily obscured, we focus on edge-on viewing
angles below. In this case, the exact spectrum of the disk+wind model 
depends on the modelling of the spatial extent of the disk and wind.
For example, if the disk
has a smaller angular extent ($\theta_0$, corresponding to a smaller covering factor), 
the emission of the inner part of the wind 
will be less absorbed by the disk. The same is true if the wind has a larger 
spatial extent (a larger $R^w_{out}$). As an approximation, for edge-on 
angles, the resulting spectrum of the 
disk+wind model can be treated as the sum of the emission from the disk and 
the wind material above the disk. Therefore, for 
simplicity, we will present a parameter study of the wind component only and do not
considering the disk. The effects of different configurations of a clumpy disk/torus have been 
studied in the literatures \citep[e.g.][]{Liu14,Tan19}.

\begin{figure*}
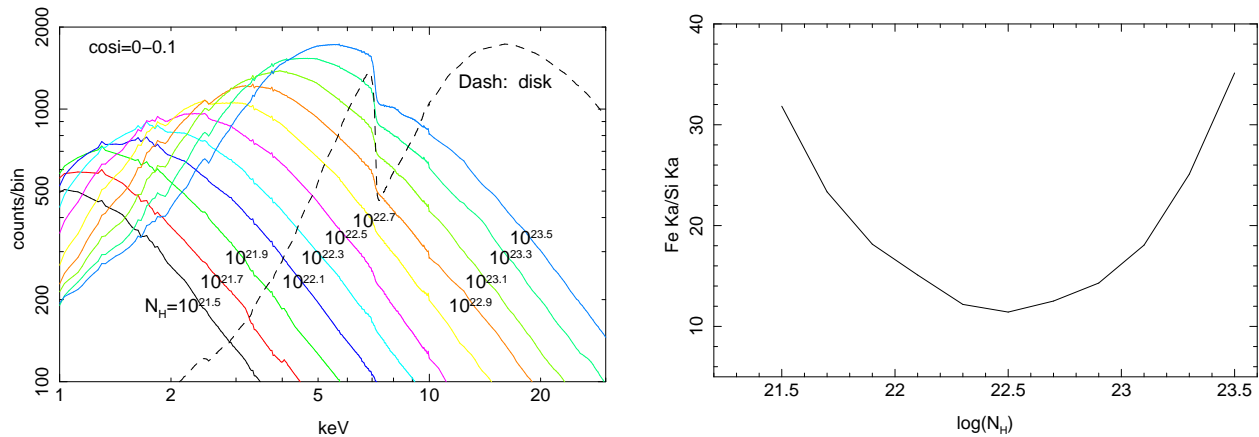

\includegraphics[width=3.4in]{NHspe.ps}
\includegraphics[width=3.4in]{NHfluo.ps}
\caption{Left: scattered X-ray continuum of
the wind model in edge-on angles with different wind column densities.
For comparison, the scattered continuum of the disk model is also plotted as 
the dashed line. Right: estimated \FeSi ratios of different wind column densities
by adding line fluxes from the disk component.
}
\end{figure*}

First, to study the effect of the material extent of the polar wind, 
we simulate the wind model with different wind column densities 
between $10^{21.5}$ and $10^{23.5}$\,cm$^{-2}$ by changing the clump density.
The results of edge-on viewing angles 
are plotted in the left panel of Figure 4. 
As can be seen, the scattered high energy emission increases with the 
column density of the wind, while the emission at low energy end decreases with 
the column density of the wind, due to absorption. As a result, the peak
of the scattered emission increases with the column density of the wind.
We note that even with a wind column density as low as $10^{21.5}$\,cm$^{-2}$, the wind
component dominates over the disk component for energies below 3 keV.
The emission of the wind is comparable to that of the disk around 5-7 keV
for a wind column density $\sim10^{23}$\,cm$^{-2}$. While for a wind column 
density around $10^{23.5}$\,cm$^{-2}$, the wind starts to dominate 
for energies within $7-10$ keV. These results clearly show that the wind component
can contribute significantly to the X-ray spectrum of type II AGN.

\begin{figure*}
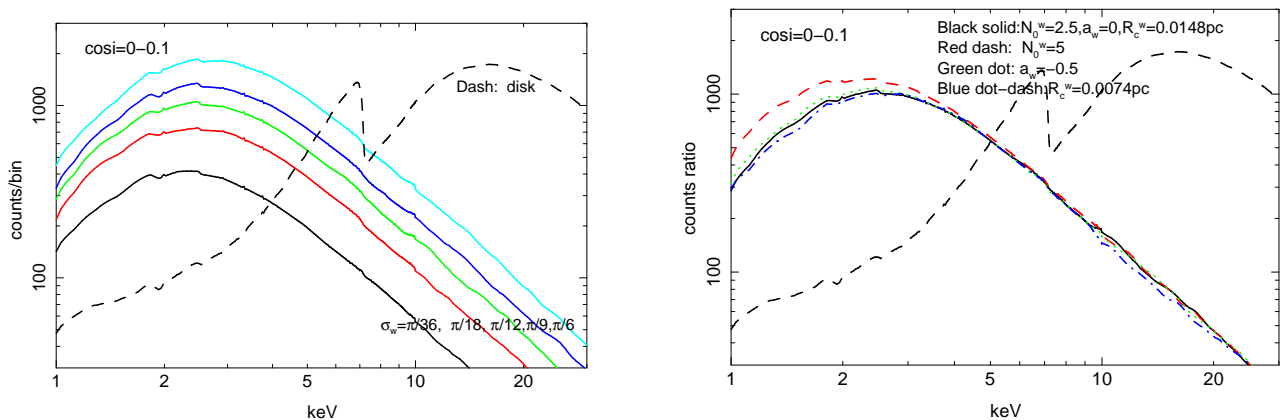

\includegraphics[width=3.5in]{Wsig_sca.ps}
\includegraphics[width=3.4in]{Clump_sca.ps}
\caption{Scattered X-ray continuum of
	 the wind model in edge-on angles with different angular extents (left), and 
	 different clump number, clump radial distribution, and clump radius (right).
For comparison, the scattered continuum of the disk model is also plotted as the black dashed line.
}
\end{figure*}

To estimate the effect of different wind column densities on
\FeSi ratios, we added the Fe \Ka and Si \Ka fluxes from
the disk model to those of the wind model of different wind 
column densities.
The result is plotted in the right panel of Figure 4. As can be seen, 
the \FeSi ratios first decrease with the wind column density within 
$10^{21.5}-10^{22.5}$\,cm$^{-2}$, reach a minimum ratio ($\sim$10) around $10^{22.5}$\,cm$^{-2}$,
and then increase with the wind column density within $10^{22.5}-10^{23.5}$\,cm$^{-2}$. 
A column density of $10^{22.5}$\,cm$^{-2}$ corresponds to an optical depth $\sim1$ for 
photons with energy around the Si K$\alpha$ region. A higher column density will result in more 
Si \Ka photons when the optical 
depth is smaller than 1 and will produce more absorption when the optical depth is above 1. 
This behavior is also shown in the scattered continuum, which is highest around 
$N_H=10^{22.5}$\,cm$^{-2}$ at energies $\sim$1.8 keV, as seen in the left panel of Figure 4.
Overall, the resulting \FeSi ratios are within $10-35$, only slightly depending on
the wind column density.

Second, to study the effect of the angular extent of the polar wind, 
we adopt different angular widths of the wind, with 
$\sigma_w=\pi/36$,$\pi/18$, $\pi/12$, $\pi/9$, and $\pi/6$. 
The results are shown in the left panel of Figure 5.
All the other parameters are kept as the default.
As can be seen, the scattered X-ray emission increases with
the angular extent of the wind. The \FeSi ratios also increase
with the angular extent. While for energies below 3 keV, 
the effect of the stronger absorption caused by the larger angular extent 
is also seen.

Finally, to study the effect of the clumpiness of the wind,
we simulate one case with a higher clump number of $N^w_0=5$. 
The clump density ($n^w_c$) is 
reduced by a factor of 2 to keep the wind column density 
the same as the default value. 
The result is plotted as the red dash line in the right panel of Figure 5. 
It looks quite similar to the default case, except below 3 keV, where
the scattered emission is a little higher than that of the default
case. We also simulate the wind model with a wind distribution index $a^w=-0.5$
and a smaller clump radius $R_c^w=0.0074$\,pc.
The results are also plotted in the right panel of Figure 5. These two parameters hardly affect 
the scattered emission.

\begin{figure*}
\includegraphics[width=3.4in]{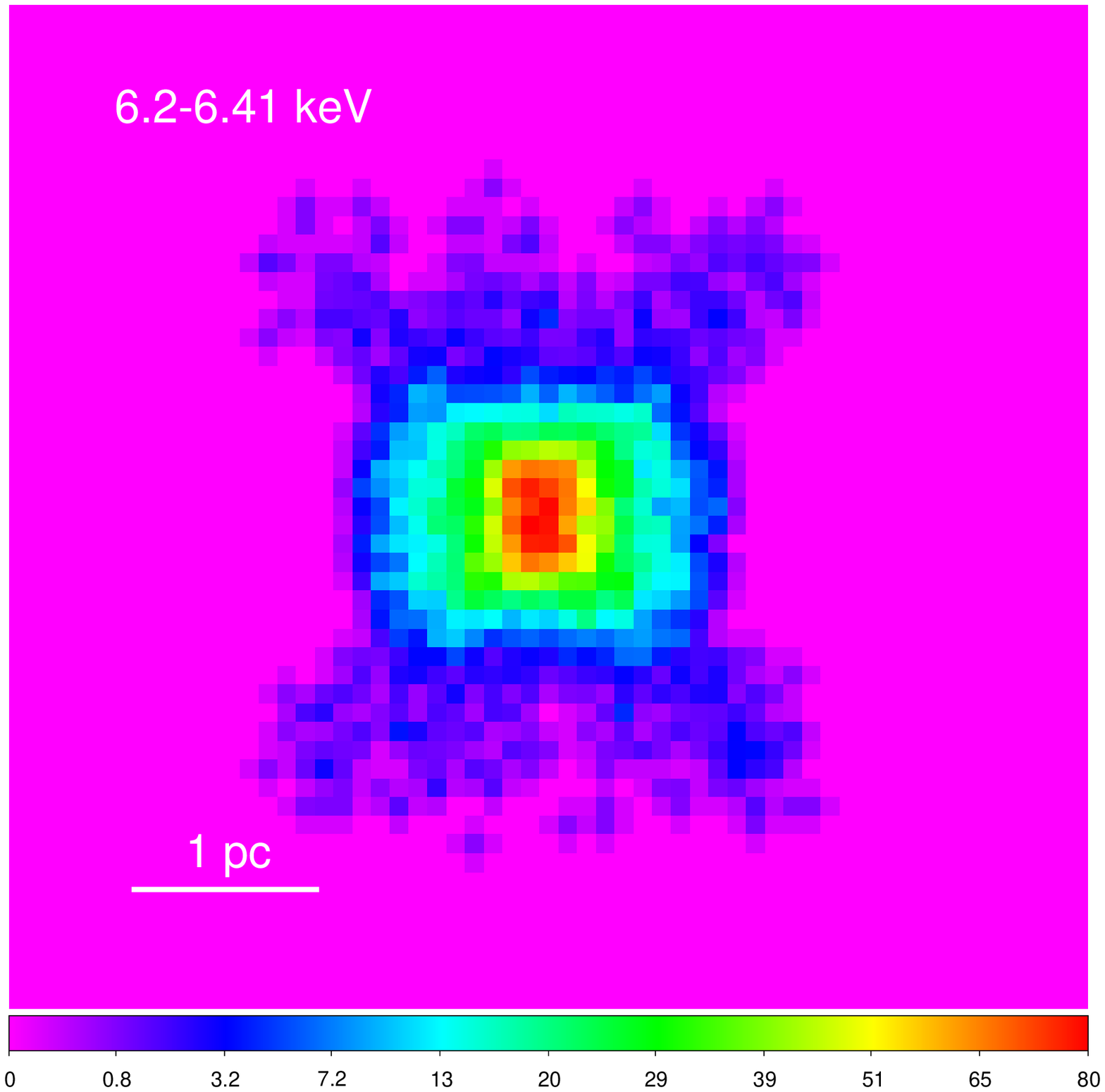}
\includegraphics[width=3.4in]{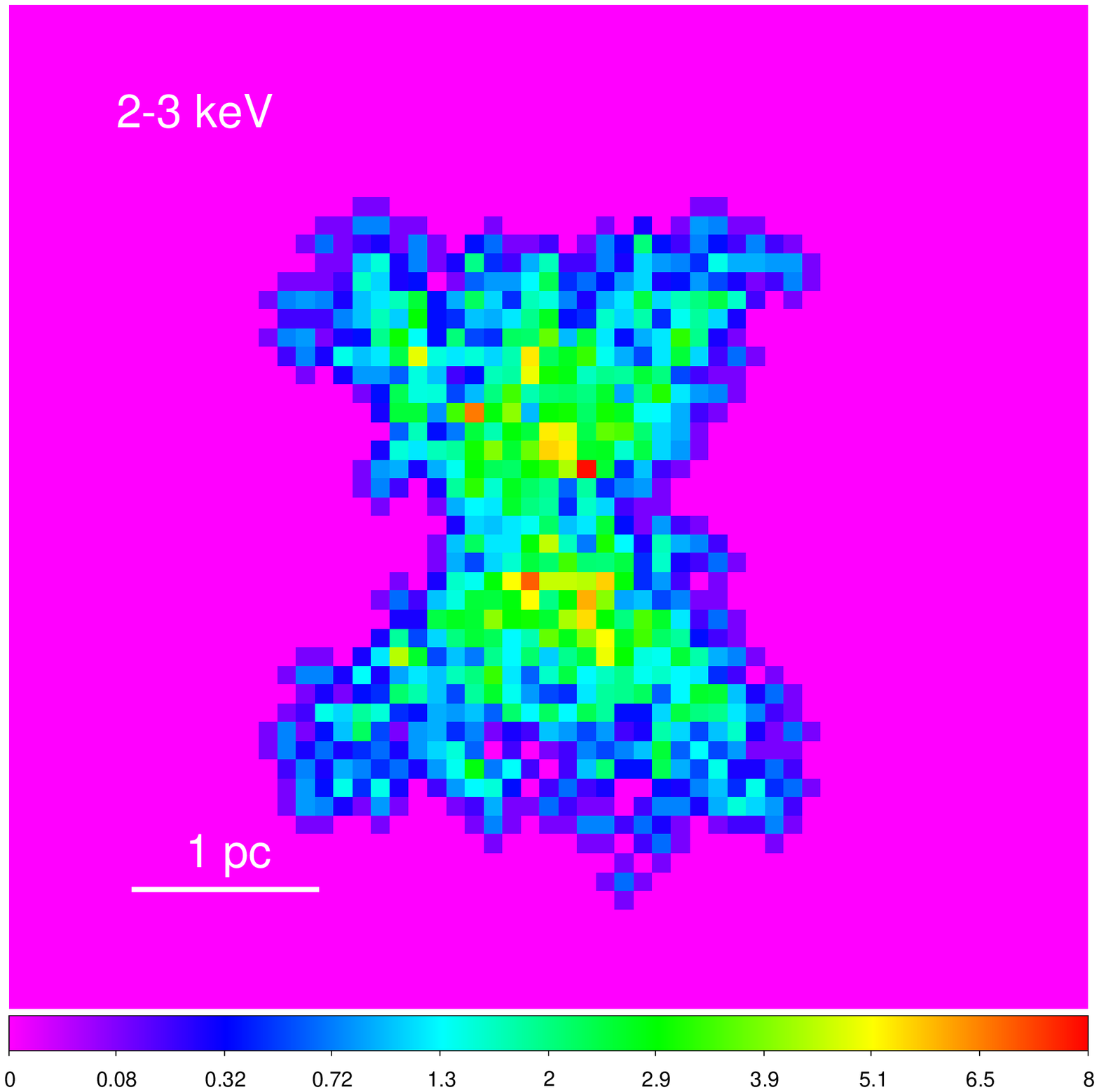}
\caption{Simulated Fe \Ka and $2-3$ keV maps of the disk+wind model extracted for photons with
	inclination angles of $\cos i<0.05$. They are constructed by projecting the photons 
	to a plane through the disk axis.
	The vertical axis is aligned with the disk axis and the horizontal axis is 
	the equatorial direction. The AGN is located at the center of the 
	image. The pixel size is 0.1 pc and a squared root color scheme is used. 
}
\end{figure*}

\subsection{Morphological results}

As shown in \S 3.1, when viewing the disk+wind model in edge-on angles, 
the wind and disk components dominate
the low-energy ($1-5$ keV) and high-energy ($>5$ keV) parts of the scattered X-ray emission, 
respectively.
This can also be seen in the morphology of the emission at different energies. In Figure 6 we 
plot the simulated maps of Fe \Ka and $2-3$ keV photons with
inclination angles of $\cos i<0.05$.
These simulated maps are plotted in physical scale, not in angular scale,
just intended to show the different morphologies of the scattered emission from different components.
The real angular extent of the polar-gas-scattered X-ray emission should be similar
to those revealed by mid-infrared interferometry observations. 
As can be seen, the inner part of the Fe \Ka map follows the
elliptical contours of the equatorial disk component; while the $2-3$ keV map is 
bipolar and traces the hollow conic wind component. 
That is, the polar-gas-scattered 
X-ray emission has an elongation similar to the emission-line-dominated soft X-ray emission
produced in the ionization cone.
This reflects the fact that the low-energy photons 
are dominated by the scattered emission of the polar wind. 
The polar wind also produce Fe \Ka photons, but with a level much fainter than those from the disk. 
It is interesting to note that 
the largest angular size of the polar winds revealed by mid-infrared interferometry ($\sim0.2''$ for
the Circinus galaxy) is 
similar to the sub-pixel spatial resolution of {\it Chandra}. We will discuss the observational
evidence of polar-gas-scattered X-ray emission in next section.

\section{Observational evidences}
\begin{figure*}
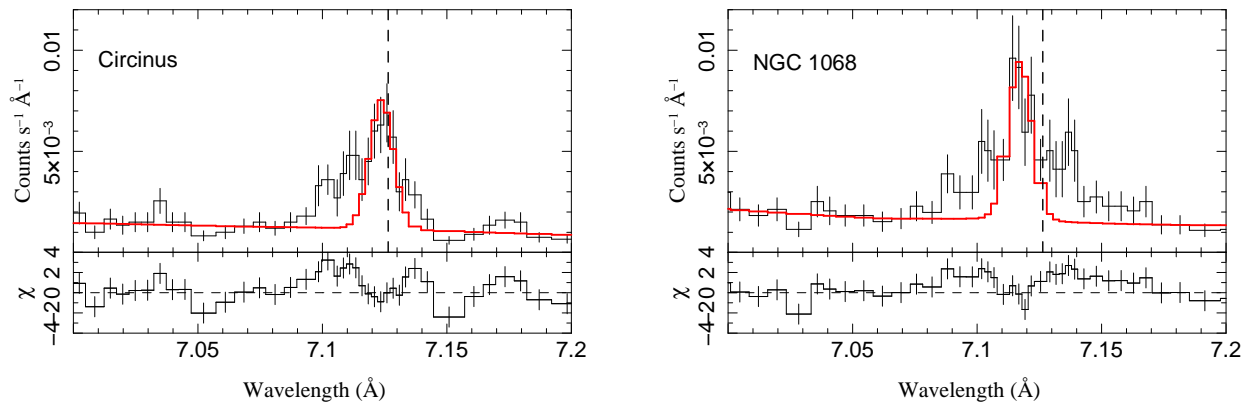

\includegraphics[width=3.4in]{LcirSi.ps}
\includegraphics[width=3.4in]{L1068Si.ps}
\caption{
\cha spectra of the Si \Ka line for the Circinus galaxy and NGC 1068, 
corrected for the systemic redshifts of the host galaxies.
The fitted Gaussian line plus a linear continuum are over-plotted as red lines,
while the vertical dashed lines indicate the rest energies of the Si \Ka line (1.73978 keV), 
which is weighted from Si K$\alpha_1$ (1.73998 keV) 
and K$\alpha_2$ (1.73938 keV).
}
\end{figure*}

\subsection{Spectral evidences}

As shown in previous section, the polar dusty gas produces strong X-ray fluorescence lines, 
such as the Si \Ka line at 1.74 keV, for edge-on viewing angles. If the dusty gas is in an 
outflowing-wind mode, 
the fluorescence lines 
will show blue-shifts with respect to the systemic velocities of the host galaxies.
The receding side of the wind is likely to be obscured, as indicated by the one-side
ionization cones and the one-side polar dust emission for the Circinus galaxy
\citep{Tri14,Sta17} and NGC 1068 \citep{LG14}.
In the soft X-ray band, \cha High Energy Transmission Grating
\citep[HETG,][]{Can05} provides the best currently available spectral resolution
of 0.012 \AA (full width half maximum, FWHM), and the line centroids can be measured 
with an accuracy $\sim$0.1 times of FWHM ($\sim0.0012$ \AA), corresponding to a velocity shift of 
$\sim50$ km\,s$^{-1}$ \citep{Ish06}. 

The Si \Ka lines of the Circinus galaxy and NGC 1068 have 
been studied with \cha HETG observations in 
the literature \citep[e.g][]{Sam01,Ogl03,Kal14,Liu16A}, but no detailed redshifts were compared with 
the host galaxies. For convenience, we replot the \cha HETG spectra of the Si \Ka line 
of the Circinus galaxy and NGC 1068, corrected for the systemic redshifts
of the host galaxies (0.00145 and 0.00379, respectively), in Figure 7.
All the archive \cha HETG data are used. The data are reprocessed following the standard process
with {\sc tgcat} script\footnote{http://tgcat.mit.edu}\citep{TG}, except 
that the zero-order centroids are determined from photons 
within $3-8$ keV, to avoid the possible contamination by off-center soft X-ray emission. 
(We noted that the Si \Ka lines presented in \citet{Liu16A} were corrected for the redshifts
of the Si \Ka lines, not the systemic redshifts of the host galaxies.)
As can be seen in Figure 7, the Si \Ka peaks are blue-shifted with respect to the systemic velocities 
of the host galaxies, although the photon statistics are limited (the 
photon counts around the Si \Ka peaks are about 80 and 50, for the Circinus galaxy and NGC 1068, 
respectively). If we fit a narrow Gaussian line (with a fixed width of 0.003 \AA) to the 
Si \Ka peaks, we obtain redshifts of $0.0011\pm0.0002$ and $0.0025\pm0.0002$, for 
the Circinus galaxy and NGC 1068, respectively (for a confidence level of 90\%). 
They correspond to outflow velocities of
$100\pm60$ and $390\pm60$ km\,s$^{-1}$, respectively. The residuals around 7.11 \A are likely due 
to the \MgXII\,\Lyb line, while those around 7.14 \A are less obvious. These results 
are consistent with an outflowing scenario of the Si K$\alpha$-emitting gas. 

\begin{figure*}
\includegraphics[width=3.4in]{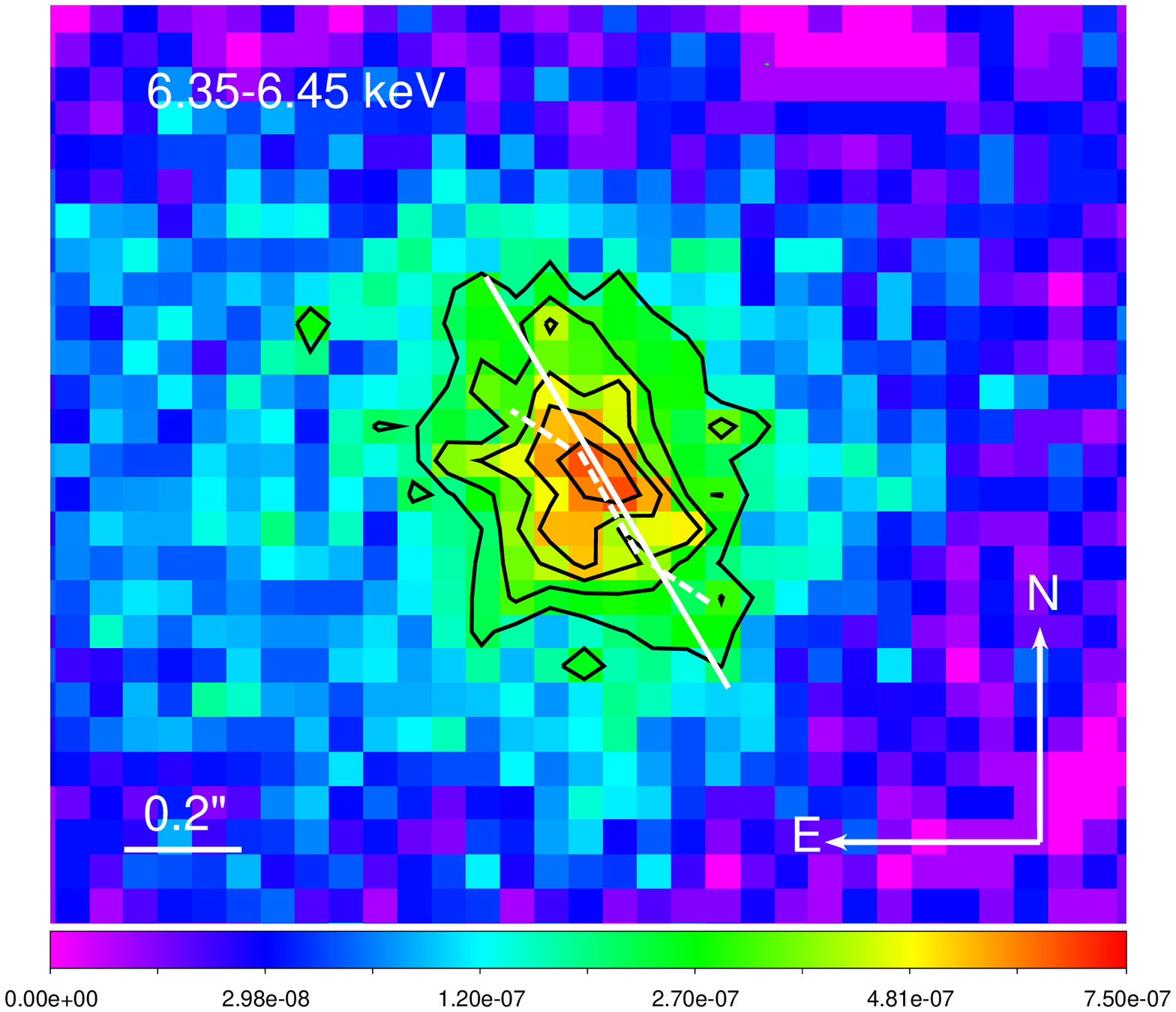}
\includegraphics[width=3.4in]{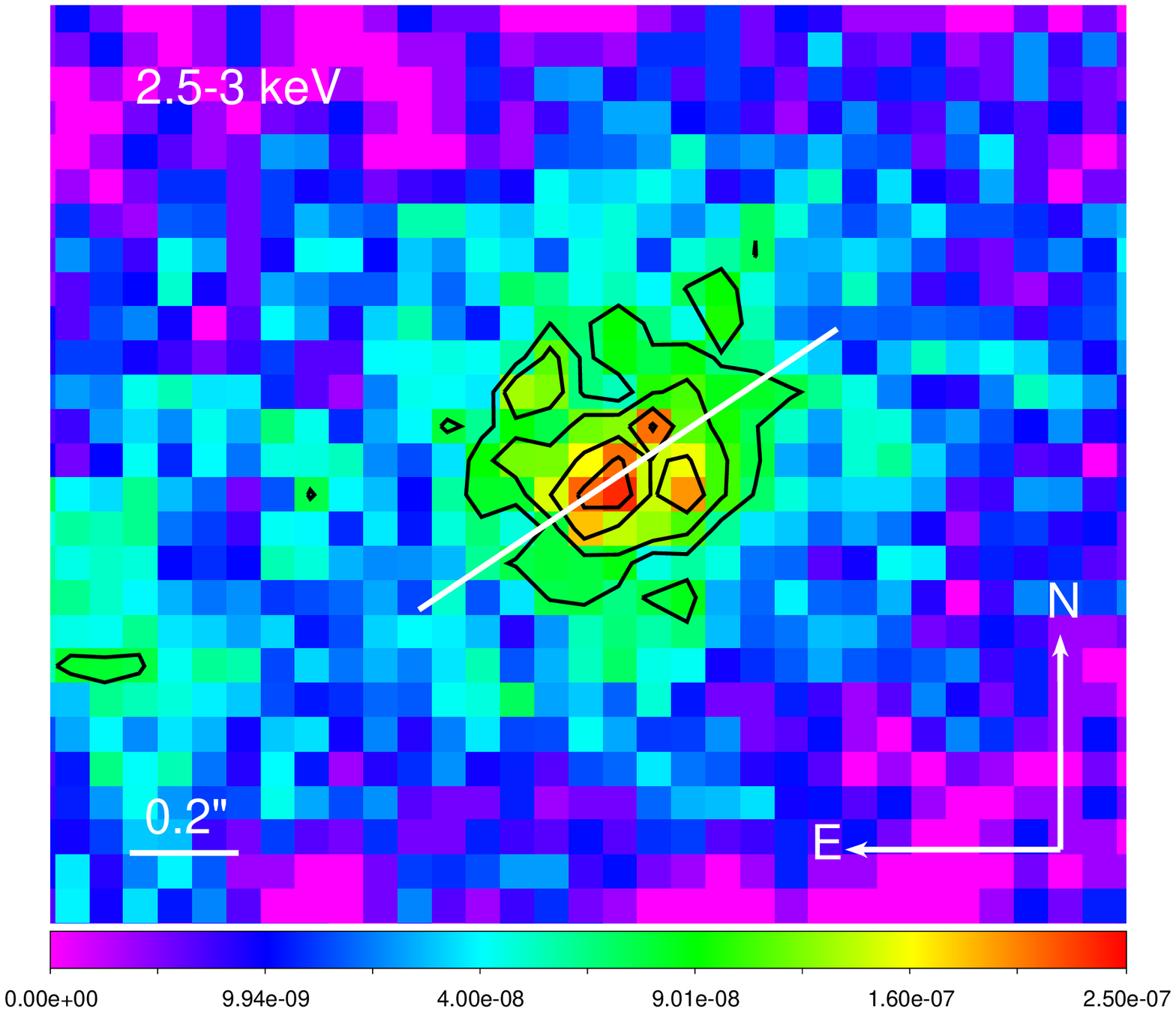}
\caption{\cha Fe \Ka and $2.5-3$ keV maps of the Circinus galaxy extracted from ObsID 12813. 
	The images are binned with a pixel size of 0.0625 arcsec. 
	The white solid lines indicate the orientations measured from the intensity distribution
	with the brightest 25 pixels, while the dashed line shows the elongation of the maser disk 
	(with an outer radius of 0.02 arcsec, not to scale).
	The Fe \Ka map is elongated 
	along the disk direction, while the $2.5-3$ keV map is perpendicular to 
	the disk.
	It indicates that the Fe \Ka and $2.5-3$ keV photons arise from different spatial regions.} 
\end{figure*}

\subsection{Morphological evidences}

With a sub-pixel spatial resolution $\sim0.2''$ \citep{Li04}, \cha can directly reveal the circumnuclear region 
of nearby Compton-thick AGN through scattered hard continuum ($2-6$ keV) and Fe \Ka lines \citep{Fab19a}.
Flattened hard continuum and Fe \Ka lines ($\sim$200 pc), perpendicular to the ionization cone, 
were found in NGC 4945 \citep[][]{Mar12,Mar17}. \citet{Mar13} found a clumpy Fe \Ka 
distribution in the Circinus galaxy. A clumpy material distribution was also revealed in NGC 1068 \citep{Mar16}.
Double-peaked Fe \Ka lines (separated by 
36 pc), both of which displaced from the centroid of $4-6$ keV photons, were discovered in 
ESO 428-G014 \citep[][]{Fab19}. Especially interesting is that an elongated Fe \Ka feature
($\sim65$ pc, but not in the $3-6$ keV continuum map), similar to
that of the CO(2-1) ALMA image, was discovered in NGC 5643, and both the Fe \Ka and 
the CO lines could be originated from the obscuring torus \citep{Fab18}.
These results are consistent with the existence of thick obscuring gas along 
the equatorial plane. 

Since the largest angular size of the polar dust probed by mid-infrared
interferometry is similar to the sub-pixel spatial resolution of {\it Chandra},
it is interesting to check the elongation of low-energy X-ray emission
for bright nearby sources.
For this purpose, we have extracted the Fe \Ka and $2.5-3$ keV images of the Circinus galaxy from the 
deepest \cha ACIS observation (ObsID 12823, 150 ks).
We also extracted the images of NGC 1068, and found their photon statistics 
(with an ACIS exposure $\sim50$ ks) are much less than
the Circinus galaxy, and thus not presented here.
The range of $2.5-3$ keV is chosen to avoid possible line contamination produced in the ionization cone
and to be dominated by the scattered emission as shown in the spectral fitting by \citet{Are14}.
The resulting images are shown in Figure 8 along with the elongation of the warped disk 
observed with the water maser \citep{Gre03}.
We measured the orientation of the maps using the intensity distribution with the brightest 25 
pixels \citep[with \textit{find\_galaxy} program written by][]{Cap02}. 
The position angles of the Fe \Ka and $2.5-3$ keV maps 
are $30^{\circ}$ and $125^{\circ}$ (with respect to the vertical direction clockwise), respectively.
They are shown as the while solid lines in Figure 8.
It is interesting to see that the elongation of the Fe \Ka map is along the direction of 
the maser disk, while the $2.5-3$ keV map is perpendicular to the maser disk. 
These behaviors are consistent with the simulated morphologies of the disk+wind model shown in \S 3.2.

We note that the elongation of the Fe \Ka emission of the Circinus galaxy on scales 
$\sim1.5''$ is more close to the horizontal direction, as already illustrated 
in \citet{SW01}. This direction and size are similar to those of the prominent 12 $\mu$m bar revealed 
in VLT/VISIR image of Circinus \citep{Sta17}. They modelled the 12 $\mu$m bar as the dusty cone edge,
which becomes bright due to the illumination of anisotropy radiation from a tilted disk.
That is, the Fe \Ka emission of the Circinus galaxy on scales $\sim1.5''$ could be the 
emission from the dusty wind.
It illustrates the complex circumnuclear region of the Circinus galaxy and a possible close 
connection between the Fe \Ka emission and the mid-infrared emission.

\section{Discussion and conclusion}

The polar dusty gas of AGN revealed by mid-IR interferometry observations can 
produce unique features in the scattered X-ray emission. 
When viewed with edge-on angles, the low-energy continuum and fluorescence photons 
produced by the polar dusty gas can
be much stronger than those originating from the equatorial disk gas 
for obscured AGN. The exact contribution of the polar gas is sensitive to
the column density ($N_H$) and the angular extent ($\sigma_w$) of the polar gas.
Different from those from the ionized gas, the scattered emission from the dusty gas 
produces self-absorption and neutral-like fluorescence lines, which are potentially a unique probe
 of the kinematics of the polar dusty gas.

As mentioned in the introduction, 
the observed \FeSi ratios are around $5-45$ \citep{Liu16}. (The Circinus galaxy 
has a largest \FeSi ratio of 60, but its Si \Ka photons are likely to be
absorbed by a dust lane on a larger scale \citep[e.g.][]{Roc06,Mez16}, 
and its real \FeSi ratio should be $\sim$40\% of the observed value.)
These \FeSi ratios are 5-10 times lower than those predicted by clumpy torus models.
The polar dusty gas 
provides a natural explanation of the observed bright Si \Ka lines.
Indeed, the predicted \FeSi ratio of the simulated disk+wind model is $\sim20$, 
close to the observed values. In turn, the bright Si \Ka lines observed in these AGN
are a strong support for the existence of the polar dusty gas in these AGN.

We examined the \cha HETG data of the Si \Ka line of the Circinus galaxy and NGC 1068, which 
show blue-shifts with respect to the systemic velocities of the host galaxies.
The results indicate that the Si K$\alpha$-emitting gas is outflowing
from the host galaxies and support the idea that the observed polar dust
is associated with a dusty wind \citep[e.g.][]{Hon12,Vol18}.
The current measurements are limited by available photon statistics, which could be 
much improved with future high throughput X-ray missions, such as Athena \citep{Nan13} and Arcus \citep{Arcus}.
The microcalorimeters of Athena will have an effective area about 100 times larger than 
\cha HETG within a larger energy range of 0.2--12 keV, with a spectral resolution of 2.5 eV. 
The proposed grating spectrometer of Arcus will have a spectral resolution of 3000 within 8--50 \AA, with 
an effective area $\sim300$ cm$^2$, more than 10 times that of \cha HETG.
Thus, X-ray spectroscopy of the fluorescence lines from the polar dusty gas is an ideal objective
for future X-ray missions, and a better understanding of the physical nature of the 
polar dusty gas is expected.

Similar to low-energy fluorescence photons, the polar dusty gas of AGN 
also produces stronger low-energy continuum compared with the disk gas. 
For type II AGN, such a polar-gas-scattered continuum could be a major contribution
to the observed spectrum. Observationally, it is not as easy to tell 
as fluorescence lines, since the 
observed X-ray continuum is also mixed with other components,
such as scattered emission from ionized gas \citep[e.g.][]{GB07}.
However, to measure accurately the properties of the disk/torus component 
of AGN, the polar-gas-scattered component must be taken into account.
As a result, constraints on the polar dusty gas might be obtained. 

The scattered X-ray emission of the polar dusty gas also produces a unique morphology.
It shows a bipolar-like feature, different from that of high-energy photons 
originating from the disk. We checked the \cha $2.5-3$ keV image of the Circinus galaxy
and found that it is elongated along the polar direction, consistent with the 
polar-gas-scattered X-ray emission. In contrast, the Fe \Ka map of the inner part of the
Circinus galaxy is elongated along the disk direction, consistent with an equatorial
disk origin. This peculiar morphology of polar-gas-scattered emission can also be better 
studied with future X-ray missions with high spatial resolution, like Lynx, 
the X-ray imager of which will have a spatial resolution of 
$0.3''$ with a much larger effective area \citep{Lynx}.

In our simulation, we have assumed that all atoms are in gas phase and not included
the dust grains, as in most X-ray scattering simulations. Different from Compton-scattering 
by atoms in gas phase, the dust grains scatter X-ray photons through small angles \citep{Ove65}.
Because the absorption cross-section of dust grains dominates over the scattering cross-section 
below 3 keV \citep[e.g.][]{Cor16}, the dust grains are unlikely to affect our results qualitatively.
Above 5 keV, the 
scattering cross-section of dust grains is comparative to the absorption cross-section, 
and the forward scattering of dust is likely to affect the scattered X-ray emission of AGN \citep{GB15}.
We plan to implement the dust grains in RefleX to investigate their effects on the scattered 
high-energy X-ray emission of AGN.
The polar wind (either dusty neutral or ionized) could affect the polarization of the observed X-ray
emission of AGN \citep[e.g.][]{DK11,Mar18a,Mar18b}. \citet{Mar18a} found that for type II AGN,
 the low-energy polarization spectrum is 
dominated by the scattered emission from the polar wind, and a neutral polar wind will produce 
unpolarized fluorescence lines imprinted on the polarized continuum. Thus the polar dusty gas
will also leave unique features in the polarization spectrum of type II AGN.
On the other hand, since the equatorial disk and the polar wind have different spatial scales, 
they may response differently to the variability of the intrinsic radiation of central AGN.
The times lags on timescale of ten years between the low-energy X-ray emission and Fe \Ka line
would be an indication of different emitting regions.

In summary, the polar dusty gas can contribute significantly 
to the observed X-ray emission of type II AGN. The polar-gas-scattered 
fluorescence lines are potentially a powerful probe of the kinematics of the polar dusty gas, 
which are crucial to understand the physical nature of the polar dusty gas.
The Si \Ka and Fe \Ka lines can be easily measured for a relatively 
large samples, and
the sources with smaller \FeSi ratios would be ideal targets for mid-IR interferometry 
follow-up observations. Future X-ray missions (such as Athena) would 
enable a large sample of obscured AGN that can be studied with 
polar-gas-scattered X-ray emission.

\section*{Acknowledgements}

We thank our referee for suggestions of polarization and variability features of the polar 
dusty wind and other comments that greatly improve the paper.  
JL is supported by National Natural Science Foundation of China (11773035 and 11811530630).
JL and SFH acknowledge support by the European Union though European Research Council Starting Grant
ERC-StG-677117 DUST-IN-THE-WIND.
CR is supported by the CONICYT+PAI Convocatoria Nacional subvencion a instalacion en la 
academia convocatoria a\~{n}o 2017 PAI77170080.
This research used data obtained from the \cha Data Archive.

\bibliographystyle{mn2e}

\begin{thebibliography}{}

\bibitem[Antonucci(1993)]{Ant93}
Antonucci, R. 1993, ARA\&A, 31, 473	

\bibitem[Antonucci et al.(1994)]{Ant94}
Antonucci, R., Hurt, T., Miller, J. 1994, ApJ, 430, 210

\bibitem[Anders \& Grevesse(1989)]{AG89}
Anders, E. \& Grevesse, N. 1989, Geochimica et Cosmochimica Acta, 53, 197

\bibitem[Asmus et al.(2016)]{Asm16}
Asmus, D., H$\ddot{\rm o}$nig, S. F., Gandhi, P. 2016, ApJ, 822, 109	

\bibitem[Ar{\'e}valo et al.(2014)]{Are14}
 Ar{\'e}valo, P., Bauer, F. E., Puccetti, S., Walton, D. J., Koss, M., Boggs, S. E., Brandt, W. N., Brightman, M.,
 Christensen, F. E., Comastri, A. et al. 2014, ApJ, 791, 81	

\bibitem[Beckert \& Duschl(2004)]{BD04}
Beckert, T.\& Duschl, W. J. 2004, A\&A, 426, 445

\bibitem[Burtscher et al.(2016)]{Bur16}
	Burtscher, L., H$\ddot{\rm o}$nig, S. F., Jaffe, W., Kishimoto, M., Lopez-Gonzaga, N., 
	Meisenheimer, K., Tristam, Konrad R. W. 2016, SPIE, 9907

\bibitem[Burtscher et al.(2013)]{Bur13}
Burtscher, L., Meisenheimer, K., Tristram, K. R. W. 2013, A\&A, 558, 149

\bibitem[Canizares et al.(2005)]{Can05}
Canizares, C. R., Davis, J. E., Dewey, D., et al. 2005, PASP, 117, 1144	

\bibitem[Cappellari(2002)]{Cap02}
Cappellari, M. 2002, MNRAS, 333, 400

\bibitem[Chan \& Krolik(2017)]{CK17}
Chan, C. \& Krolik, J. H. 2017, ApJ, 843, 58 

\bibitem[Comastri(2004)]{Com04}
Comastri, A. 2004, ASSL, 308, 245

\bibitem[Corrales et al.(2016)]{Cor16}
Corrales, L. R., Garc{\'i}a, J., Wilms, J. Baganoff, F. 2016, MNRAS, 458, 1345	

\bibitem[Dorodnitsyn et al. (2008)]{Dor08}
Dorodnitsyn, A., Kallman, T., Proga, D. 2008, ApJ, 675, 5

\bibitem[Dorodnitsyn \& Kallman(2011)]{DK11}
Dorodnitsyn, A. \& Kallman, T. 2011, Ap\&SS, 336, 245

\bibitem[Dorodnistyn et al.(2016)]{Dor16}
Dorodnitsyn, A., Kallman, T., Proga, D. 2016, ApJ, 819, 115 



\bibitem[Fabbiano et al.(2018)]{Fab18}
Fabbiano, G., Paggi, A., Siemiginowska, A., Elvis, M. 2018, ApJ, 869, L36

\bibitem[Fabbiano et al.(2019)]{Fab19}
Fabbiano, G., Siemiginowska, A., Paggi, A., Elvis, M., Volonteri, M., Mayer, L., Karovska, M., Maksym, W. P., Risaliti, G., Wang, J. 2019, ApJ, 870, 69

\bibitem[Fabbiano(2019)]{Fab19a}
Fabbiano, G. 2019, arXiv:190301970

\bibitem[George \& Fabian(1991)]{GF91}
George, I. M. \& Fabian, A. C. 1991, MNRAS, 249, 352

\bibitem[Gohil \& Ballantyne(2015)]{GB15}
Gohil, R. \& Ballantyne, D. R. 2015, MNRAS, 449, 1449


\bibitem[Greenhill et al.(2003)]{Gre03}
Greenhill, L. J., Booth, R. S., Ellingsen, S. P., Herrnstein, J. R., Jauncey, D. L., 
McCulloch, P. M., Moran, J. M., Norris, R. P., Reynolds, J. E., Tzioumis, A. K.
 2003, ApJ, 590, 162	

\bibitem[Guainazzi \& Bianchi(2007)]{GB07}
Guainazzi, M. \& Bianchi, S. 2007, MNRAS, 374, 1290 

\bibitem[H$\ddot{\rm o}$nig et al.(2012)]{Hon12}
H$\ddot{\rm o}$nig, S. F., Kishimoto, M., Antonucci, R., Marconi, A., Prieto, M. A., Tristram, K., Weigelt, G. 2012, ApJ, 755, 149	

\bibitem[H$\ddot{\rm o}$nig et al.(2013)]{Hon13}
H$\ddot{\rm o}$nig, S. F., Kishimoto, M., Tristram, K. R. W., Prieto, M. A., Gandhi, P., Asmus, D., Antonucci, R., Burtscher, L., Duschl, W. J., Weigelt, G. 2013, ApJ, 771, 87	

\bibitem[H$\ddot{\rm o}$nig \& Kishimoto(2017)]{HK10}
H$\ddot{\rm o}$nig, S. F., Kishimoto, M. 2010, A\&A, 523, 27	


\bibitem[H$\ddot{\rm o}$nig \& Kishimoto(2017)]{HK17}
H$\ddot{\rm o}$nig, S. F., Kishimoto, M. 2017, ApJ, 838, L20	

\bibitem[Huenemoerder et al.(2011)]{TG}
Huenemoerder D.P., et al., 2011, AJ, 141, 129

\bibitem[Ishibashi et al.(2006)]{Ish06}
Ishibashi, K., Dewey, D., Huenemoerder, D. P., Testa, P. 2006, 644, L117

\bibitem[Jaffe et al.(2004)]{Jaf04}
Jaffe, W., Meisenheimer, K., R$\ddot{\rm o}$ttgering, H. J. A., Leinert, Ch., Richichi, A., Chesneau, O.,
Fraix-Burnet, D., Glazenborg-Kluttig, A., Granato, G.-L., Graser, U., et al. 
 2004, Nature, 429, 47	

\bibitem[Kallman et al.(2014)]{Kal14}
Kallman, T., Evans, D. A., Marshall, H., Canizares, C., Longinotti, A., Nowak, M.,
Schulz, N. 2014, ApJ, 780, 121

\bibitem[Kishimoto et al.(2007)]{Kis07}
Kishimoto, M., H$\ddot{\rm o}$nig, S. F., Beckert, T., Weigelt, G. 2007, A\&A, 476, 713


\bibitem[Krolik \& Kriss(1995)]{KK95}
Krolik, J. H. \& Kriss, G. A. 1995, ApJ, 447, 512 

\bibitem[Lee et al.(2001)]{Lee01}
Lee, J. et al. 2001, ApJ, 554, L13

\bibitem[Leftley et al.(2018)]{Lef18}
	Leftley, J. H., Tristram, K. R. W., H$\ddot{\rm o}$nig, S. F., Kishimoto, M., Asmus, D., Gandhi,
			P. 2018, ApJ, 862, 17	

\bibitem[Li et al.(2004)]{Li04}
Li, J., Kastner, Joel H., Prigozhin, G. Y., Schulz, N. S., Feigelson, E. D., 
Getman, K. V. 2004, ApJ, 610, 1204	

\bibitem[Liu et al.(2016)]{Liu16}
Liu, J., Liu, Y., Li, X., Xu, W., Gou, L., Cheng, C. 2016, MNRAS, 459, L100	

\bibitem[Liu (2016)]{Liu16A}
Liu, J. 2016, MNRAS, 459, L105 

\bibitem[Liu \& Li(2014)]{Liu14}
Liu, Y. \& Li, X. 2014, ApJ, 787, 52

\bibitem[L{\'o}pez-Gonzaga et al.(2014)]{LG14}
	L{\'o}pez-Gonzaga, N., Jaffe, W., Burtscher, L., Tristram, K. R. W., Meisenheimer, K.
	 2014, A\&A, 565, 71	

\bibitem[L{\'o}pez-Gonzaga et al.(2016)]{LG16}
	L{\'o}pez-Gonzaga, N., Burtscher, L., Tristram, K. R. W., Meisenheimer, K., Schartmann, M. 
	2016, A\&A, 591, 47

\bibitem[Lyu \& Rieke(2018)]{Lyu18} 
Lyu, J. \& Rieke, G. H. 2018, ApJ, 866, 92	

\bibitem[Lynx team(2018)]{Lynx}
The Lynx Team, 2018, arXiv:1809.09642

\bibitem[Magdziarz \& Zdziarski(1995)]{MZ95}
Murphy, K. D. \& Yaqoob, T. 2009, MNRAS, 397, 1549	


\bibitem[Marinucci et al.(2012)]{Mar12} 
Marinucci, A., Risaliti, G., Wang, J., Nardini, E., Elvis, M., Fabbiano, G., Bianchi, S., 
Matt, G. 2012, MNRAS, 423, L6

\bibitem[Marinucci et al.(2013)]{Mar13} 
Marinucci, A.; Miniutti, G.; Bianchi, S.; Matt, G.; Risaliti, G. 2013, MNRAS, 436, 2500

\bibitem[Marinucci et al.(2016)]{Mar16} 
Marinucci, A., Bianchi, S., Matt, G. et al. 2016, MNRAS, 456, L94

\bibitem[Marinucci et al.(2017)]{Mar17} 
Marinucci, A.; Bianchi, S.; Fabbiano, G.; Matt, G.; Risaliti, G.; Nardini, E.; Wang, J. 
2017, MNRAS, 470, 4039

%



\bibitem[Marin et al.(2018a)]{Mar18a} 
Marin, F., Dov$\check{c}$iak, M., Muleri, F., Kislat, F. F., Krawczynski, H. S. 2018, MNRAS, 473, 1286


\bibitem[Marin et al.(2018b)]{Mar18b} 
Marin, F., Dov$\check{c}$iak, M., Kammoun, E. S. MNRAS, 478, 950



\bibitem[Mehdipour \& Costantini(2018)]{MC18} 
Mehdipour, M. \& Costantini, E. 2018, A\&A, 619, 20 

\bibitem[Mezcua et al.(2016)]{Mez16} 
	Mezcua, M., Prieto, M. A., Fern{\'a}ndez-Ontiveros, J. A., Tristram, K. R. W.
 2016, MNRAS, 457, L94	

\bibitem[Murphy \& Yaqoob(2009)]{MY09}
Murphy, K. D. \& Yaqoob, T. 2009, MNRAS, 397, 1549	


\bibitem[Nandra \& Pounds(1994)]{NP94}
Nandra, K. \& Pounds, K. A. 1994, MNRAS, 268, 405

\bibitem[Nandra et al.(2013)]{Nan13}
Nandra, K. et al. 2013, arXiv:1306.2307

\bibitem[Nenkova et al.(2008a)]{Nen08a}
Nenkova, M., Sirocky, M.~M., Nikutta, R., Ivezi{\'c}, {\v Z}., \& Elitzur, M.\ 2008, ApJ, 685, 147

\bibitem[Nenkova et al.(2008b)]{Nen08b}
Nenkova, M., Sirocky, M.~M., Nikutta, R., Ivezi{\'c}, {\v Z}., \& Elitzur, M.\ 2008, ApJ, 685, 160

\bibitem[Netzer(2015)]{Net15}
Netzer, H. 2015, ARA\&A, 53, 365	

\bibitem[Ogle et al.(2003)]{Ogl03}
Ogle, P. M., Brookings, T., Canizares, C. R., Lee, J. C., Marshall, H. L. 2003, A\&A, 402, 849O

\bibitem[Overbeck(1965)]{Ove65}
Overbeck, J. W. 1965, ApJ, 141, 864O	
%
%

\bibitem[Paltani \& Ricci(2017)]{PR17}
Paltani, S.; Ricci, C. 2017, A\&A, 607, 31	

\bibitem[Proga et al.(2000)]{Pro00}
Proga, D., Stone, J. M., Kallman, T. R. 2000, ApJ, 543, 686

%
\bibitem[Ricci et al.(2014)]{Ric14}
Ricci, C., Ueda, Y., Paltani, S., Ichikawa, K., Gandhi, P., Awaki, H. 2014, MNRAS, 441, 3622

\bibitem[Roche et al.(2006)]{Roc06}
	Roche, P. F., Packham, C., Telesco, C. M., Radomski, J. T., Alonso-Herrero, A.,
			Aitken, D. K., Colina, L.; Perlman, E. 2006, MNRAS, 367, 1689	

\bibitem[Sambruna et al.(2001)]{Sam01}
Sambruna, R. M., Netzer, H., Kaspi, S., Brandt, W. N., Chartas, G., Garmire, G. P.,
Nousek, J. A., Weaver, K. A. 2001, ApJ, 546, L13

\bibitem[Smith \& Wilson(2001)]{SW01}
Smith, D. A. \& Wilson, A. S. 2001, ApJ, 557, 180

\bibitem[Smith et al.(2016)]{Arcus}
Smith, R. K. et al. 2016, SPIE, 9905, 4

\bibitem[Stalevski et al.(2017)]{Sta17}
Stalevski, M., Asmus, D., Tristram, K. R. W. 2017, MNRAS, 472, 3854	

\bibitem[Stalevski et al.(2017)]{Sta19}
Stalevski, M., Tristram, K. R. W., Asmus, D. 2019, MNRAS, 484, 3334 

\bibitem[Tanimoto et al.(2019)]{Tan19}
Tanimoto, A., Ueda, Y., Odaka, H., Kawaguchi, T., Fukazawa, Y., 
Kawamuro, T. 2019, ApJ, 877, 95

\bibitem[Tristram et al.(2014)]{Tri14}
	Tristram, K. R. W., Burtscher, L., Jaffe, W., Meisenheimer, K., H$\ddot{\rm o}$nig, S. F.,
			Kishimoto, M., Schartmann, M., Weigelt, G. 2014, A\&A, 563, 82	
\bibitem[Vollmer et al.(2018)]{Vol18}
Vollmer, B., Schartmann, M., Burtscher, L., Marin, F., H$\ddot{\rm o}$nig, S., Davies, R., 
Goosmann, R. 2018, A\&A, 615, 164

\bibitem[Williamson et al.(2019)]{Wil19}
Williamson, D., Venanzi, M., H$\ddot{\rm o}$nig, S. F. 2019, ApJ, 876, 137




\end{thebibliography}

\end{document}